\documentclass[%
 pre,reprint,
groupedaddress,
 amsmath,amssymb,
 aps,
longbibliography
]{revtex4-1}

\usepackage{graphicx}
\usepackage{dcolumn}
\usepackage{bm}
\usepackage{float}

\usepackage{hyperref}

\usepackage[usenames, dvipsnames]{xcolor}

\hypersetup{
	colorlinks,
	citecolor=blue,      
	filecolor=red,
	linkcolor=blue,
	urlcolor=blue,
	hyperfigures
}

\begin{document}

\title{Self-induced hydrodynamic coil-stretch transition of active polymers}

\author{Achal Mahajan}
\author{David Saintillan}\email[Electronic address: ]{dstn@ucsd.edu}
\affiliation{Department of Mechanical and Aerospace Engineering, University of California San Diego, 9500 Gilman Drive, La Jolla, CA 92093, USA}


\begin{abstract}
We analyze the conformational dynamics and statistical properties of an active polymer model. The polymer is described as a freely-jointed bead-rod chain subject to stochastic active force dipoles that act on the suspending solvent where they drive long-ranged fluid flows. Using Langevin simulations of isolated chains in unconfined domains, we show how the coupling of active flows with polymer conformations leads to emergent dynamics. Systems with contractile dipoles behave similarly to passive Brownian chains with enhanced fluctuations due to dipolar flows. In systems with extensile dipoles, however, our simulations uncover an active coil-stretch transition whereby the polymer spontaneously unfolds and stretches out in its own self-induced hydrodynamic flow, and we characterize this transition in terms of a dimensionless activity parameter comparing active dipolar forces to thermal fluctuations.\ We discuss our findings in the context of the classic coil-stretch transition of passive polymers in extensional flows, and complement our simulations with a simple kinetic model for an active trimer.
\end{abstract}

\maketitle

\section{Introduction}

Biological polymers and other filamentous structures are involved in a myriad of complex dynamical processes essential for the functioning of the cell. In many cases, these polymers are subject to out-of-equilibrium microscopic stresses driven by ATP-powered active processes \cite{kolomeisky2007molecular}, resulting in unexpected emergent properties such as self-organization and spontaneous motion. Examples of such systems abound and include: the coordinated action of dynein motors along the microtubule bundle comprising eukariotic cilia and flagella, where active sliding forces exerted by the motors result in spontaneous beating and locomotion \cite{RHHJ2010,BS2019}; the unidirectional transport of cargo by kinesin and dynein motors along tracks of microtubules in the cell cytoskeleton \cite{VKS2011,RRVC2018}; the contraction of the actin cortex under the action of myosin motors during various processes such as cytokinesis \cite{ATO2019} and platelet formation \cite{BBG2020}; and the generation of cytoplasmic streaming flows by dynein motors marching along microtubules during oocyte development \cite{GWPG2012,SDLSG2021}. These biological polymers and molecular motors have also been extracted and utilized in vitro, for instance in motility assays \cite{schaller2010polar} or as building blocks for active materials with various self-emergent properties \cite{sanchez2011cilia,sanchez2012spontaneous,ND2017}. 

Recent experiments by Zidovska \textit{et al.}\ \cite{zidovska2013micron} have demonstrated that the interior of the cell nucleus, which encloses chromatin, the functional form of DNA in cells, also displays coherent motions indicative of ATP-powered activity. While the detailed microscopic mechanisms responsible for these motions are still under debate, there is supporting evidence that they result from stresses exerted by active enzymes such as RNA polymerases, helicases and topoisomerases. In a minimal hydrodynamic description \cite{bruinsma2014chromatin}, these active processes can be idealized as exerting force dipoles on chromatin chain sections, which are transmitted to the nucleoplasm by viscous friction, thereby driving long-ranged fluid flows.\ Based on this concept, Saintillan \textit{et al.}\ \cite{saintillan2018extensile} performed numerical simulations  of a long flexible polymer chain confined in a spherical cavity and subject to active force dipoles distributed stochastically along the polymer and applied to the viscous solvent. They found that, in the case of extensile dipoles, the resulting flows in the nucleoplasm conspire to align chromatin chain segments, thereby triggering a feedback loop that results in large-scale coherent motions by a mechanism similar to the generic instability of extensile active nematics \cite{SS2008b,SS2013,Gao15a,doostmohammadi2018active}. 

A wide variety of other theoretical and computational models for active polymers have been proposed in recent years \cite{WG2020}, guided by some of the experimental systems discussed above. Some of these models have focused on semiflexible polymer chains, actuated by random correlated velocities \cite{LPG2020} or composed of connected active Brownian particles (ABPs) \cite{LA2015,EGW2016,MEGW2019}, either with or without hydrodynamic interactions.  Other models have considered free-draining flexible Rouse chains, either composed of ABPs \cite{BLM2018,AS2020}, subjected to correlated fluctuating forces \cite{O2018} or dipoles \cite{PSV2019}, or to active isotropic extensile or contractile forces \cite{LPBS2021}; these latter models, however, all neglected hydrodynamic interactions.

In this work, we analyze the conformational dynamics of an active flexible bead-rod linear polymer chain subjected to active force dipoles that drive fluid disturbances in the suspending solvent. The model is based on the past work of Saintillan \textit{et al.}\ \cite{saintillan2018extensile}, but we provide here a more systematic characterization of the role of activity and hydrodynamic interactions on chain dynamics and conformations in unconfined dilute systems. Our numerical simulations demonstrate that extensile dipolar activity results in the spontaneous stretching of the chains above a critical level of activity by driving a coherent long-ranged extension-dominated flow that overcomes thermal fluctuations and unravels the polymer. This `active coil--stretch transition' is reminiscent of the transition exhibited by passive flexible polymers in externally applied extensional flows \cite{D1974}, but is internally driven by dipolar activity in the present case.  The paper is organized as follows.\ First, we introduce in Sec.~\ref{sec:model} the model for the polymer dynamics and motor activity following \cite{saintillan2018extensile}. We present numerical results in Sec.~\ref{sec:results}, where we characterize the role of activity on the transition from coiled to stretched configurations, and also analyze a simple kinetic model for an active timer based on the Fokker-Planck equation.\ We summarize our results and conclude in Sec.~\ref{sec:conclusions}.

\section{Active polymer model\label{sec:model}}

\subsection{Langevin formulation}

\begin{figure}[t]
    \centering
        \includegraphics[width=0.95\linewidth]{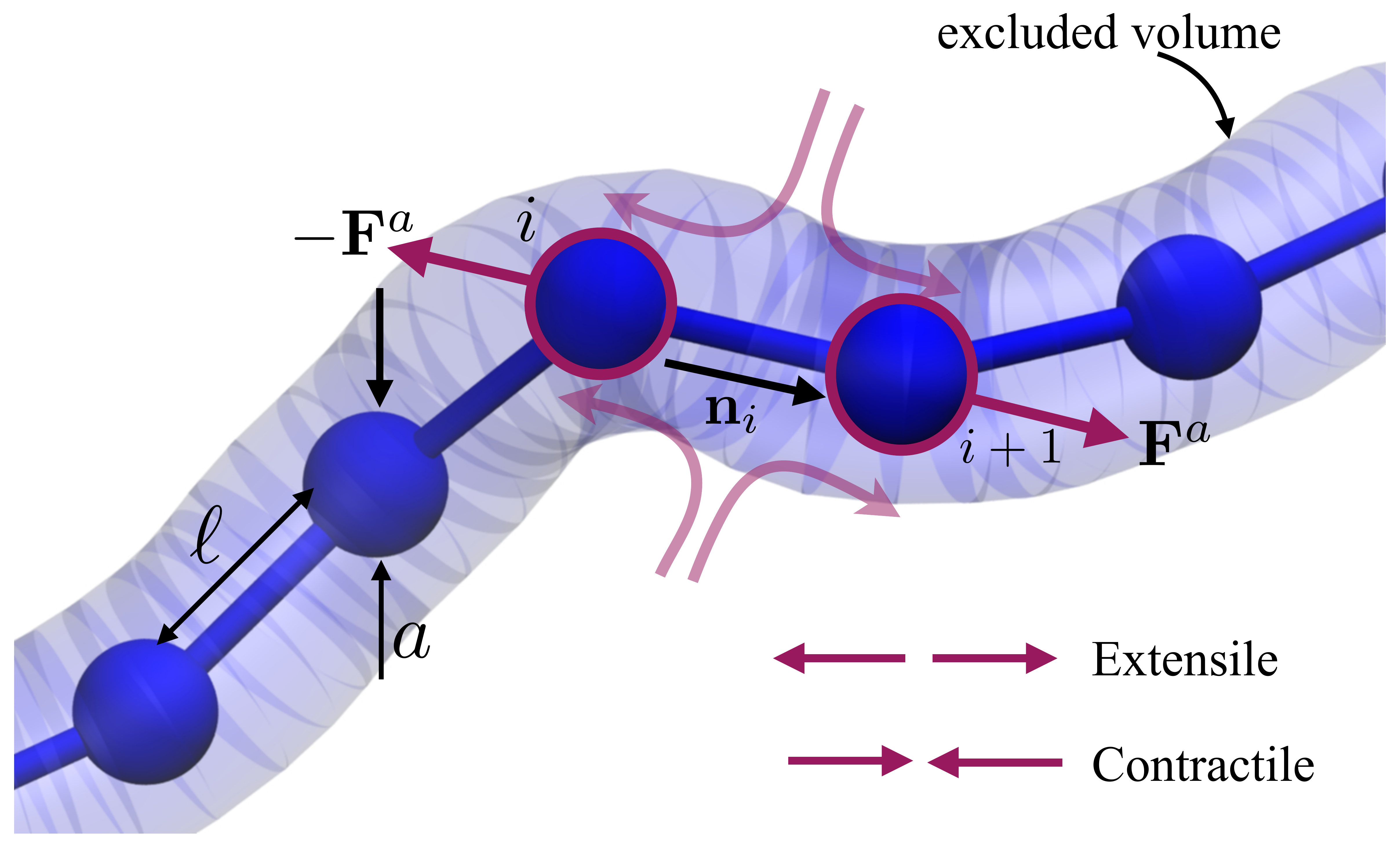}
    \caption{Coarse-grained active polymer: The polymer is modeled as a long freely-jointed chain of $N$ beads with hydrodynamic radius $a$ and connected by rigid links of length $\ell$. Active dipoles (extensile or contractile) bind stochastically to individual links, with forces applied directly on the suspending fluid.}
   \label{Fig:Fig1}
\end{figure}

We study the dynamics and conformations of active flexible polymer chains in an unconfined environment. Figure~\ref{Fig:Fig1} shows a graphical representation of the polymer model, which consists of a linear chain of $N$ beads of hydrodynamic radius $a$ connected by $N-1$ freely jointed inextensible rods of length $\ell$ with no bending resistance or rotational constraints. The chain is subject to tension forces, thermal fluctuations and excluded volume interactions. In addition, we model activity in terms of force dipoles (pairs of equal and opposite forces) that are applied to the solvent and drive hydrodynamic disturbances as we explain in Sec.~\ref{sec:activef} \cite{bruinsma2014chromatin,saintillan2018extensile}. 
We denote by $\mathbf{r}_i(t)$ the position of bead $i$, and by $\mathbf{n}_i(t)=(\mathbf{r}_{i+1}-\mathbf{r}_i)/\ell$ the unit vector pointing from bead $i$ to bead $i+1$. At low Reynolds number, the motion of the chain is overdamped and satisfies the Langevin equation~\cite{ottinger1996stochastic}:
\begin{equation}
\frac{\mathrm{d}\mathbf{r}_i}{\mathrm{d}t}=\mathbf{u}^a(\mathbf{r}_i,t)+\frac{1}{\zeta}\left[\mathbf{F}^c_i(t)+\mathbf{F}^e_i(t)\right]+\boldsymbol{\xi}_i(t). \label{Eq:Eq1}
\end{equation}
The first term on the right-hand side describes advection by the active fluid flow $\mathbf{u}^a$ induced by ATP-powered molecular motors, which we discuss in Sec.~\ref{sec:activef}. The second term captures motion under the internal constraint forces $\mathbf{F}^c_i$ ensuring inextensibility (see Sec.~\ref{sec:tension}) and under excluded volume forces $\mathbf{F}^e_i$ acting on bead $i$, with hydrodynamic friction coefficient $\zeta=6\pi\eta a$. Excluded volume interactions are captured by a soft repulsive potential:
\begin{equation}
\mathbf{F}_i^{e} = -\sum_{j \ne i} \nabla_i \Phi(\mathbf{r}_i - \mathbf{r}_j), \quad \text{with} \quad \Phi(\mathbf{r}) = \frac{\Phi_0}{r^n}, \label{eq:ev}
\end{equation}
with $n=3$, which is truncated whenever $|\mathbf{r}_i-\mathbf{r}_j|>\ell$. The parameter $\Phi_0$ is ad hoc and selected to ensure that the chain does not cross itself.\ Finally, the last term in Eq.~(\ref{Eq:Eq1}) captures Brownian displacements and satisfies the fluctuation--dissipation theorem:
\begin{equation}
 \langle \boldsymbol{\xi}_i(t)\rangle=\mathbf{0}, \quad \langle \boldsymbol{\xi}_i(t)\boldsymbol{\xi}_j(t')\rangle=2 D_b \mathbf{I}\,\delta_{ij}\delta(t-t'), \label{Eq:flucdiss}
\end{equation}
where $D_b=k_\mathrm{B} T/\zeta$ is the Brownian diffusivity of an isolated bead. 

Note that the Langevin equation (\ref{Eq:Eq1}) models the polymer as a free-draining chain.\ Constraint and excluded volume forces are applied locally on each bead and do not drive any hydrodynamic interactions: the only flow in the solvent is assumed to be that induced by active dipoles. We note that this is an approximation that is convenient computationally. This description can be improved by accounting for hydrodynamic interactions between distant chain segments: we consider this more general case in Appendix~\ref{sec:appendix}, where we find that hydrodynamic interactions tend to damp the effect of activity due to the enhanced viscous drag on the chain.

\subsection{Active forces and hydrodynamic flow\label{sec:activef}}

A key feature of our model is the inclusion of active forces and flows, which model the effects of the activity of ATP-powered motor proteins performing work along the polymer chain. The stresses exerted by these motors are coarse-grained in the form of active dipolar force pairs that occur on the scale of one chain link and drive fluid disturbances (Fig.~\ref{Fig:Fig1}) \cite{saintillan2018extensile}. As the motors bind and unbind stochastically along the chain, we assume that each link can be either active (dipole on) or inactive (dipole off), and the time in a given state is drawn from an exponential distribution with two distinct rates $k_\mathrm{on}$ and $k_\mathrm{off}$. These rates set the probability $p_a = k_\mathrm{on} /(k_\mathrm{on} + k_\mathrm{off} )$  that any given link is active at a point in time. Note that $p_a$ can also be interpreted as the mean fraction of active links in the system.

When a link is in the active state, two equal and opposite forces are applied to the viscous surrounding fluid at the positions of the two end beads:
\begin{equation}
 \mathbf{F}^{a}_i = -F_0 \mathbf{n}_i(t), \quad \mathbf{F}^{a}_{i+1} = F_0 \mathbf{n}_i(t), \label{Eq:Eq7} 
\end{equation}
where $F_0$ is the strength of the active force, assumed to be constant. This effectively describes a force dipole with magnitude $F_0\ell$, which can be either extensile ($\leftarrow\rightarrow$) or contractile ($\rightarrow\leftarrow$) as shown in Fig.~\ref{Fig:Fig1}. In the present model, we assume that these two forces are applied directly to the solvent, where they drive an active hydrodynamic flow that can be obtained as
\begin{equation}
 \mathbf{u}^a (\mathbf{r}_i)  = \sum_{j=1}^N \mathbf{G}(\mathbf{r}_i;\mathbf{r}_j) \cdot \mathbf{F}_j^a,\label{Eq:Eq6}
\end{equation}
where $\mathbf{G}$ denotes the Oseen tensor in free space:
\begin{equation}
\mathbf{G}(\mathbf{r}_i,\mathbf{r}_j) = \frac{1}{8 \pi \mu r_{ij}} \Big[\mathbf{I} + \hat{\mathbf{r}}_{ij} \hat{\mathbf{r}}_{ij}\Big], \label{eq:oseen}
\end{equation}
where $r_{ij} = |\mathbf{r}_{i}-\mathbf{r}_{j}|$ is the distance between beads $i$ and $j$, and  $\hat{\mathbf{r}}_{ij}=(\mathbf{r}_{i}-\mathbf{r}_{j})/r_{ij}$ is the unit vector pointing from bead $j$ to bead $i$.

 {Note that, within the model described here, active stresses are only applied to the fluid and have no direct effect on chain dynamics other than via long-ranged hydrodynamic interactions. As a model for the action of ATP-powered molecular motors, this assumes that any chain reconfiguration resulting from direct forces by the motors would occur on sub-Kuhn-step length scales that are unresolved by our model. The present description could also serve as a model for a freely-jointed chain of rigid bimetallic autophoretic rods \cite{PBKWMS2006} that drive flows via phoretic slip layers.  }

\subsection{Constraint forces and numerical algorithm \label{sec:tension}}

Chain inextensibility is imposed by means of the constraint forces $\mathbf{F}^c_i$ in Eq.~(\ref{Eq:Eq1}), which are calculated to ensure that the length of each link remains constant and equal to $\ell$. Specifically, we express these constraint forces in terms of scalar tensions $T_i$, which are Lagrange multipliers: 
\begin{equation}
\mathbf{F}_i^{c} = T_i \mathbf{n}_i - T_{i-1} \mathbf{n}_{i-1}.\label{Eq:Eq10}
\end{equation}
The two ends of the polymer are force-free, which reflects in the boundary conditions $T_0 = T_{N } = 0$. To solve for the tensions, we employ the algorithm of Liu \cite{liu1989flexible}, which uses a semi-implicit predictor-corrector scheme to ensure that the length of each link is preserved at the end of every time step. The first step in the time-marching scheme is an unconstrained explicit Euler step for the bead positions, in which the tension forces are omitted:
\begin{equation}
\tilde{\mathbf{r}}_i^{n+1} = \mathbf{r}_i^{n} + \Delta t \left [ \mathbf{u}^a (\mathbf{r}_i^{n}) + \boldsymbol{\xi}_{i}^n + \zeta^{-1}\mathbf{F}_{i,n}^{ev}\right ]. \label{Eq:Eq11}
\end{equation}
The second step corrects the position $\tilde{\mathbf{r}}_i^{n+1}$ to account for the tensions, which are evaluated implicitly at time $t_{n+1}$:
\begin{equation}
\mathbf{r}_i^{n+1} = \tilde{\mathbf{r}}_i^{n+1} + \zeta^{-1} \Delta t \left [T_i^{n+1} \mathbf{n}_i^n - T_{i-1}^{n+1} \mathbf{n}_{i-1}^n \right ]. \label{Eq:Eq12}
\end{equation}
Eq.~(\ref{Eq:Eq12}) can be further recast in terms of the unit directors between beads as
\begin{equation}
\mathbf{n}_i^{n+1}\! =\! \tilde{\mathbf{n}}_i^{n+1}\! +\! \frac{\Delta t}{\zeta \ell}\! \left [T_{i+1}^{n+1} \mathbf{n}_{i+1}^n\! -\! 2T_{i}^{n+1}\mathbf{n}_{i}^n\! +\! T_{i-1}^{n+1} \mathbf{n}_{i-1}^n \right ]. \label{Eq:Eq13}\vspace{0.1cm}
\end{equation}
Applying the inextensibility constraint $|\mathbf{n}_i^{n+1}|^2 = 1$ then leads to an equation for the tensions:
\begin{align}
\frac{2\Delta t}{\zeta \ell}&\!\left [T_{i+1}^{n+1} \mathbf {n}_{i+1}^n\! - 2T_{i}^{n+1}\mathbf{n}_{i}^n\! + T_{i-1}^{n+1} \mathbf{n}_{i-1}^n \right ] \cdot \tilde{\mathbf{n}}_{i}^n\! =\! 1\!-\! |\tilde{\mathbf{n}}_{i}^n|^2  \nonumber \\ 
&\!\!\!\!\! - \frac{\Delta t^2}{\zeta^2 \ell^2} \left |T_{i+1}^{n+1} \mathbf{n}_{i+1}^n - 2T_{i}^{n+1}\mathbf{n}_{i}^n + T_{i-1}^{n+1} \mathbf{n}_{i-1}^n \right |^2.\label{Eq:Eq14}
\end{align}
This system of quadratic equations can be solved iteratively, where each iteration involves inverting a linear tridiagonal system corresponding to the left-hand side in Eq.~(\ref{Eq:Eq14}). A small number of iterations (typically $\lesssim 10$) is sufficient to achieve convergence. 

\subsection{Scalings and parameters \label{sec:scaling}}

In the following, we present results in dimensionless form, where all the variables are scaled using the following characteristic time, length and force scales:
\begin{equation}
 t_c = \frac{\zeta \ell^2}{k_\mathrm{B} T}, \qquad \ell_c = \ell, \qquad F_c = \frac{k_\mathrm{B} T}{\ell}. \label{Eq:Eq5}
\end{equation}
With this choice, the dimensionless active dipole strength becomes
\begin{equation}
\sigma_0 = \frac{F_0 \ell}{k_\mathrm{B} T}, \label{Eq:Eq6}
\end{equation}
where $\sigma_0 > 0$ and $\sigma_0 < 0$ represent extensile and contractile systems, respectively. We also introduce an activity parameter $A$, defined as \begin{equation}
    A=p_a\sigma_0=\frac{k_\mathrm{on}}{k_\mathrm{on}+k_\mathrm{off}}\frac{F_0 \ell}{k_\mathrm{B} T},
\end{equation}
which can be viewed an effective active dipole strength corrected for the stochasticity of molecular motors \cite{saintillan2018extensile}.

\section{Results and discussion\label{sec:results}}

\subsection{Conformational dynamics and flow fields \label{sec:dynamics}}

\begin{figure}[t]
    \centering
    \includegraphics[width=1\linewidth]{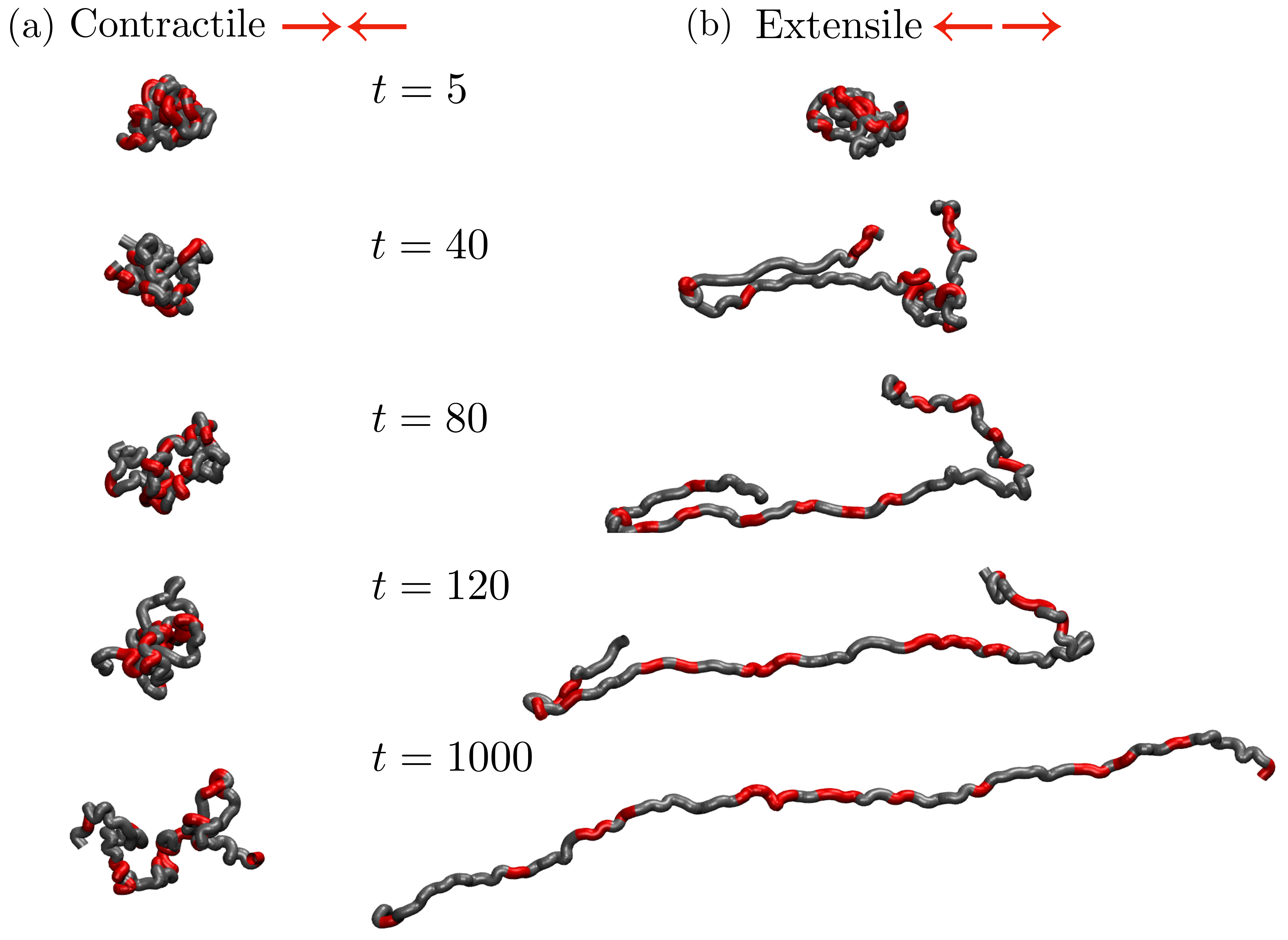}
    \caption{Snapshots from two simulations of (a) a contractile system ($A=-10$) and (b) an extensile system ($A=10$) with $N=100$, $k_{\mathrm{on}}=100$ and  $k_{\mathrm{off}}=500$. The chains were initially prepared as random walks. Red segments show the instantaneous positions of active force dipoles. The contractile chain swells but remains in an isotropic coiled state, whereas the extensile chain progressively unfolds to reach a fluctuating stretched conformation. See Supplemental Material \cite{Note1} for movies of the dynamics.}
    \label{Fig:Fig2}
\end{figure}

\begin{figure*}[t]
    \centering
    \includegraphics[width=\textwidth]{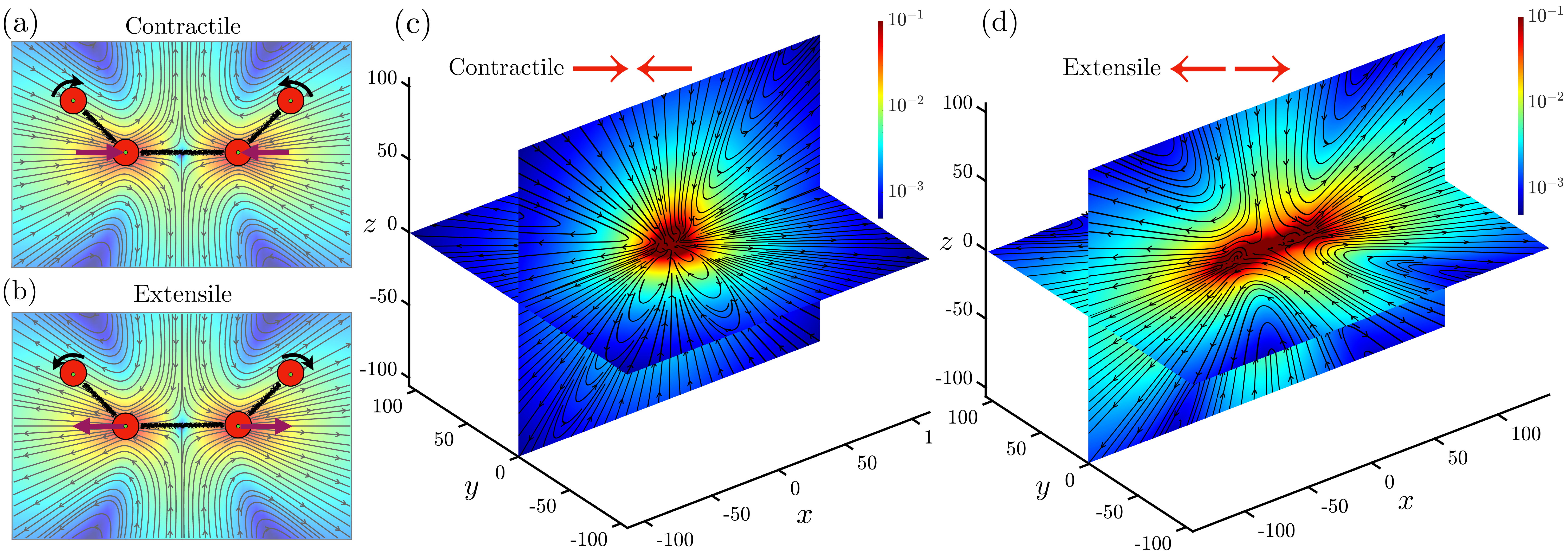}\vspace{-0.2cm}
    \caption{(a,b)  {Schematic illustrating the instantaneous} flow fields induced by a contractile (a) and an extensile (b) dipole, respectively, and their effect on chain dynamics  {in a representative case of a short chain composed of $N=4$ beads.} The purple arrows show the direction of the active forces, whereas the black arrows show the direction of rotation of the neighboring links  {due the active flow induced by the dipole.} (c,d) Time-averaged flow fields induced by (a) a contractile system ($A=-10$) and (b) an extensile system ($A=10$)  {in simulations} with $N=100$, $k_{\mathrm{on}}=100$ and  $k_{\mathrm{off}}=500$. In the contractile case, the chain is coiled at the origin  {and the flow field was calculated in a fixed coordinate system.} In the extensile case, the polymer is stretched and the coordinate system was rotated to align the principal axis of the chain with the $x$ direction.  {In both (c) and (d), the flow fields were averaged over 200 frames equally spaced in time over a total duration of 400 dimensionless time units after reaching statistical steady state.}  }\vspace{-0.1cm}
    \label{Fig:flow}
\end{figure*}

We performed simulations of polymer chains composed of $N=3-150$ beads, for varying values of $k_\mathrm{on}$, $k_\mathrm{off}$ and $\sigma_0$.\ Figure~\ref{Fig:Fig2} shows snapshots from simulations of two active chains, one contractile and one extensile, with $N=100$ and $A=\pm 10$; also see the videos in the Supplemental Material \cite{Note1}.\ Both systems were prepared as random walks at $t=0$.\ At short times, excluded volume forces cause both polymers to swell in all directions, while dipolar flows are being driven locally along the chains. In the contractile case, the chain continues to swell slightly as the flows induced by the active dipoles constantly rearrange neighboring links, with no net emergent alignment. At statistical steady-state, the contractile chain remains in an isotropic coiled state, with thermal fluctuations enhanced by the dipolar fluid flows. The case of extensile activity is markedly different.\ There, neighboring links align in the dipolar flows, which have extensional symmetry, resulting in long chain segments opening up and stretching, until the chain fully unfolds into a stretched conformation where activity-induced flow alignment competes against thermal fluctuations, which favor the coiled state. Similar dynamics had previously been observed by Saintillan \textit{et al.}\ \cite{saintillan2018extensile}.\ The mean direction of alignment of the chain is arbitrary, but remains roughly constant over the course of the transient.\ At steady state, the axis of alignment very slowly diffuses as a result of Brownian motion, which also introduces stochasticity in the flow field by causing fluctuations of the dipole orientations.

These dynamics can be understood based on the flow fields induced by individual dipoles as depicted in Fig.~\ref{Fig:flow}(a,b).\ In the contractile case shown in (a), the flow is compressional along the axis of the dipole and extensional in the perpendicular plane, which tends to rotate neighboring chain links so that they align at $\approx 90^\mathrm{o}$, thus forming kinks in the chain. Conversely, in the extensile case shown in (b), the flow is extensional along the axis of the dipole, which tends to rotate neighboring links so that they align with the dipole, thus unfolding and straightening the chain.  This flow-induced alignment observed in extensile systems not only affects nearest neighbors but is long-ranged due to the slow decay as $1/r^{2}$ of  dipolar flows in the Stokes regime. Furthermore, the effect is self-amplifying: as several chain segments carrying dipoles align with each other, their flow fields superimpose coherently to further enhance alignment. This positive feedback loop ultimately results in the full stretching of the chain under its self-induced flow. Time-averaged flows around the two polymer chains of Fig.~\ref{Fig:Fig2} are shown in Fig.~\ref{Fig:flow}(c,d). The flow in the contractile case in (c) displays extension and compression along various directions but shows no clear large-scale symmetry. On the other hand, dipolar alignment in the extensile case shown in (d) conspires to drive a self-sustained extensional flow on the scale of the polymer chain, which acts to stabilize the unfolded stretched conformation. 

The observed transition from an initial coiled state to a stretched configuration in extensile systems is reminiscent of the classic coil-stretch transition of flexible polymers in linear extensional flows \cite{D1974}, where the stretching is due to the viscous drag on the polymer by an externally applied flow. Here, the transition is instead an emergent phenomenon resulting from the spontaneous alignment of the chain under the extensile dipoles that it carries. We discuss this analogy in more detail further below.  

\subsection{Steady-state statistical properties}

\begin{figure}[b]
    \centering
    \includegraphics[width=1\linewidth]{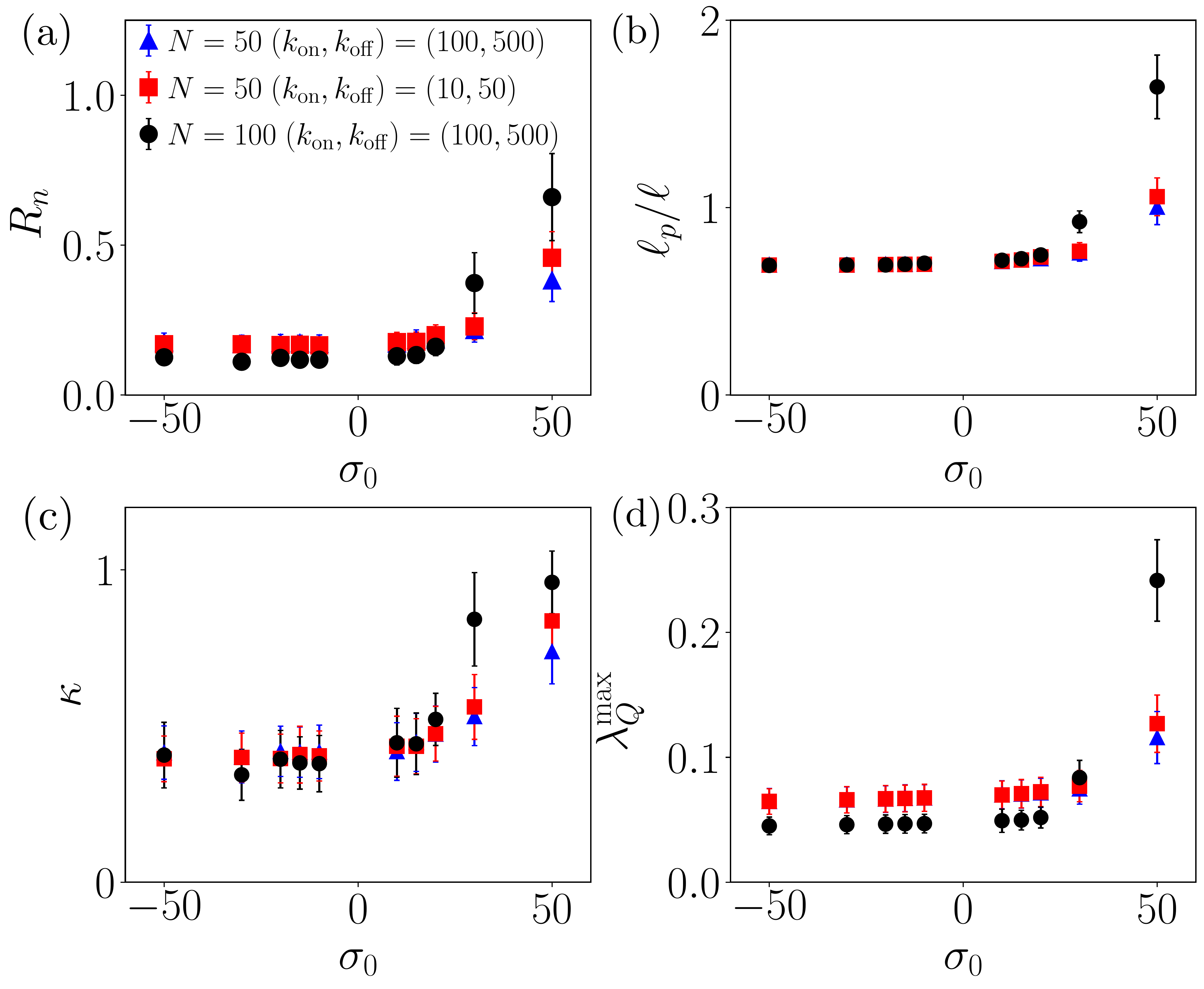}\vspace{-0.2cm}
    \caption{Average steady-state values of (a) the scaled end-to-end distance $R_n$, (b) the effective persistence length $\ell_p/\ell$, (c) the relative shape anisotropy $\kappa$, and (d) the nematic scalar order parameter $\lambda_Q^\mathrm{max}$ as functions of $\sigma_0$ for chains of length $N=50$ and $100$. In the case of $N=50$, two different combinations of on/off rates giving the same value of $p_a=1/6$ were used: $(k_\mathrm{on},k_\mathrm{off})=(100,500)$ (blue triangles) and $(10,50)$ (red squares). The simulation with $N=100$ uses $(k_\mathrm{on},k_\mathrm{off})=(100,500)$ (black circles).}
    \label{Fig:Fig3}
\end{figure}

We now discuss conformational properties at statistical steady state, which we characterize in Fig.~\ref{Fig:Fig3} showing various structural order parameters as functions of $\sigma_0$. Three sets of simulations are shown, two with $N=50$ and one with $N=100$.\ In the case of $N=50$, two different combinations of on/off rates were used: $(k_\mathrm{on},k_\mathrm{off})=(100,500)$ and $(10,50)$, which have the same value of $p_a$ and thus the same mean number of dipoles.\ All shown quantities were measured after reaching steady state and were averaged over time and over an ensemble of 5 different simulations.\ We first plot in Fig.~\ref{Fig:Fig3}(a) the relative end-to-end distance defined as 
\begin{equation}
    R_n = \frac{|\mathbf{r}_N-\mathbf{r}_1|}{N-1}, \label{Eq:Eq15}
\end{equation}
with maximum value of 1 corresponding to a perfectly straight polymer chain.\ For a perfect random walk (coiled configuration), the expected value is $\smash{R_n= \ell \sqrt{N-1}}$.\ As shown in Fig.~\ref{Fig:Fig3}(a), the end-to-end distance is unaffected by activity in the contractile case ($\sigma_0<0$), where it sligthly exceeds the random walk prediction due to excluded volume effects.\ In the extensile case, however, $R_n$ starts increasing above a critical positive value of $\sigma_0$, reaching $\approx 0.6$ for the longest chain when $\sigma_0=50$. Longer chains tend to stretch more for the same dipole strength, since they are decorated by more dipoles thus enhancing the self-unfolding effect described in Sec.~\ref{sec:dynamics}. The two simulations with $N=50$ provide insight into the role of the on/off rates on the dynamics: the end-to-end distance, as well as the other order parameters we discuss below, are only weakly affected by the actual values of  $k_\mathrm{on}$ and $k_\mathrm{off}$ at a given value of $p_a$. Stretching is slightly weaker in the case with the larger rates (blue triangles), as there is more stochasticity in the dipoles in that case. 

The unfolding and stretching of extensile chains under their self-induced flows can also be interpreted as an increase in orientational correlations along the chain, i.e., as an increase in the effective rigidity of the polymer. We quantify this using the mean persistence length, which for a bead-rod chain can be defined as \cite{koslover2013discretizing}
\begin{equation}
 \frac{\ell_p}{\ell} = \frac{1}{2}\left(\frac{1 + \xi}{1- \xi}\right), \quad \text{where} \quad \xi = \langle \mathbf{n}_i \cdot \mathbf{n}_{i+1} \rangle. \label{Eq:Eq18}
\end{equation}
As shown in Fig.~\ref{Fig:Fig3}(b), the persistence length shows similar trends as the end-to-end distance: it is independent of activity in the contractile case, but shoots up with activity in extensile systems above the transition to stretched conformations. 

\begin{figure*}
    \centering  
    \includegraphics[width=\textwidth]{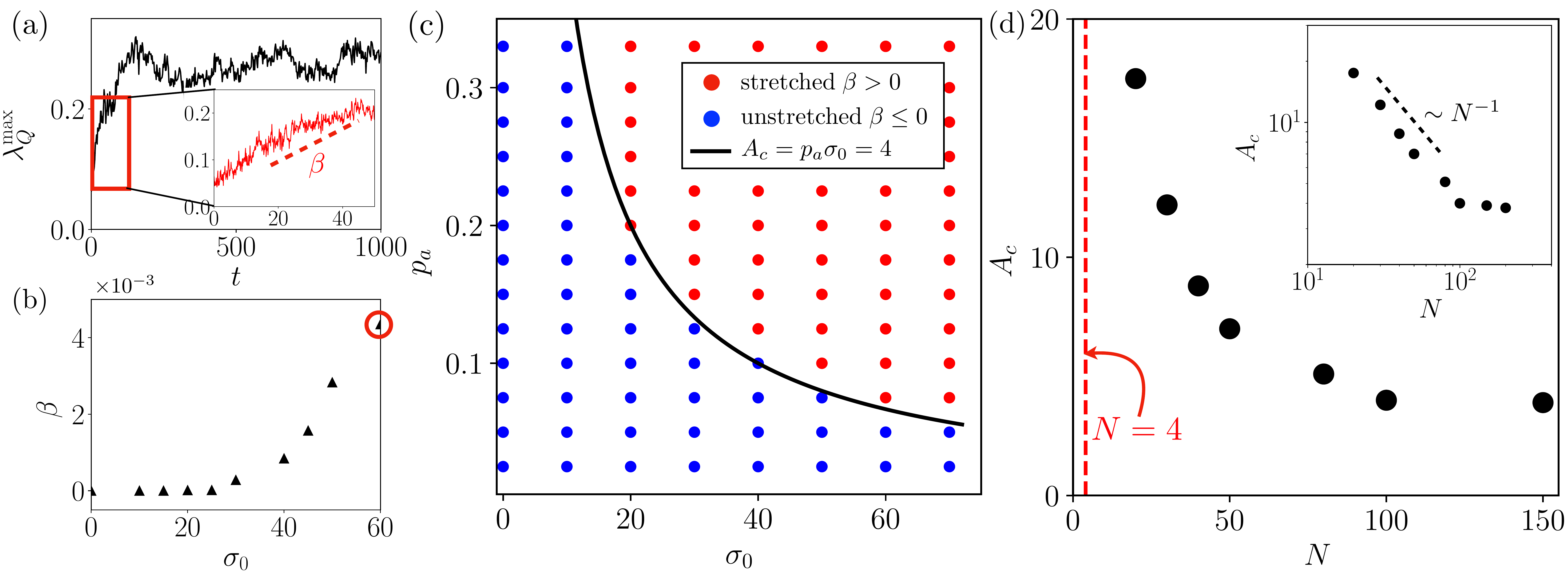}
    \caption{(a) Time evolution of the nematic scalar order parameter $\lambda_Q^\mathrm{max}$ in a simulation with $N=100$,  {$(k_\mathrm{on},k_\mathrm{off})=(100,500)$ ($p_a=1/6$)}, and $\sigma_0= 60$.  The inset shows the initial regime of linear growth, from which we define the growth rate $\beta$. (b)~Dependence of growth rate $\beta$ on dipole strength $\sigma_0$ for $N=100$ and  {$(k_\mathrm{on},k_\mathrm{off})=(100,500)$ ($p_a=1/6$)}, showing the transition to stretching above a critical value. The red circle corresponds to the simulation shown in (a). (c) Phase diagram for the active coil-stretch transition in the $(\sigma_0,p_a)$ plane. The boundary between the two domains is well captured by a curve of constant $A=A_c$ (black curve). (d) Critical activity level $A_c$ for the coil-stretch transition as a function of chain length $N$. The coil-stretch transition is only observed for chains of length $N\ge 4$. The inset shows the same data in a log-log plot, highlighting  {an apparent scaling of} $A_c\sim N^{-1}$ for intermediate chain lengths.}
    \label{Fig:Fig6}
\end{figure*}

Another measure of the mean polymer conformation is provided by the radius of gyration tensor $\mathbf{R}_G$, defined as
\begin{equation}
\mathbf{R}_G = \frac{1}{N}\sum_{i=1}^N ( \mathbf{r}_i-\mathbf{r}_{cm})(\mathbf{r}_i-\mathbf{r}_{cm}), \label{Eq:Eq16}
\end{equation}
where $\mathbf{r}_{cm}$ is the instantaneous center-of-mass position. Its eigenvalues $\alpha_1, \alpha_2$ and $\alpha_3$, characterize the extent of the polymer mass distribution along its eigendirections, with a larger eigenvalue corresponding a net extension of the chain in the corresponding direction. These eigenvalues can be used to define a shape anisotropy parameter $\kappa$ as
\begin{equation}
 \kappa = 1 - 3\frac{\alpha_1\alpha_2 + \alpha_1\alpha_3 +  \alpha_2\alpha_3}{(\alpha_1 + \alpha_2 + \alpha_3)^2}. \label{Eq:Eq17}
\end{equation}
A straight linear polymer (e.g., $\alpha_1=1$, $\alpha_2=\alpha_3=0$) has an anisotropy of $\kappa=1$, whereas an isotropic or spherically symmetric configuration ($\alpha_1=\alpha_2=\alpha_3$) has an anisotropy of $\kappa=0$. The dependence of $\kappa$ on activity is depicted in Fig.~\ref{Fig:Fig3}(c). In contractile systems, the mean value of $\kappa$ is independent of activity and relatively low at $\approx 0.3$: this non-zero value is due to the fact that instantaneous conformations on which we calculate $\kappa$ are not perfectly isotropic due to fluctuations, even though on average the system shows no preferred direction. In the presence of extensile activity, anisotropy increases as the chains extend along the principal direction of the gyration tensor, with $\kappa$ reaching nearly 1 for $N=100$ at $\sigma_0=50$.

Finally, we also characterize global nematic alignment of chain segments by plotting in Fig.~\ref{Fig:Fig3}(d) the scalar nematic order parameter $\lambda_Q^\mathrm{max}$, defined as the maximum eigenvalue  of the mean nematic order tensor 
\begin{equation}
\mathbf{Q} = \frac{1}{N - 1} \sum_{i=1}^{N-1} \left(\mathbf{n}_{i} \mathbf{n}_{i} - \frac{\mathbf{I}}{3}\right), \label{Eq:Eq19}
\end{equation}
and we note that $\lambda_Q^\mathrm{max}$ is in the range of 0 to 2/3. Once again, we observe similar trends with respect to activity, with no discernible effect of $\sigma_0$ in contractile systems, but a sharp increase of the maximum eigenvalue above the coil-stretch transition, especially in the case of long chains ($N=100$).

\subsection{Coil-stretch transition: Phase diagram}

We further characterize the dependence of the coil-stretch transition on system parameters in Fig.~\ref{Fig:Fig6}, where we consider extensile systems only. We systematically determine the location of the transition by considering the time evolution of the nematic scalar order parameter, which exhibits a positive growth only in simulations in which stretching occurs.  {Note that other quantities, such as the end-to-end distance $R_n$ or relative shape anisotropic $\kappa$, could also be used for that purpose, though we find that the nematic order parameter is particular useful as it displays growth starting from very short times in cases where stretching occurs.}   The time evolution of $\smash{\lambda_Q^\mathrm{max}}$ in one such case is plotted in Fig.~\ref{Fig:Fig6}(a), where a regime of linear growth is observed followed by a plateau. A growth rate $\beta$ can be computed during the initial regime, which is roughly zero for configurations that stay coiled but becomes positive when stretching occurs. This is illustrated in Fig.~\ref{Fig:Fig6}(b), where we plot $\beta$ as a function of $\sigma_0$ for  {a fixed choice of the on- and off-rates of $(k_\mathrm{on},k_\mathrm{off})=(100,500)$, for which $p_a=1/6$. Note that, as already observed in Fig.~\ref{Fig:Fig3}, it is the value of $p_a$ that governs the transition, rather than the intrinsic values of $k_\mathrm{on}$ and $k_\mathrm{off}$.} The trends are similar to those found in Fig.~\ref{Fig:Fig3}, with a clear transition to stretching occurring at a critical dipole strength of $\sigma_0\approx 30$. Similar simulations were performed by varying both $\sigma_0$ and $p_a$, and the results are summarized in a phase diagram in Fig.~\ref{Fig:Fig6}(c) in the case $N=100$, where we highlight  parameter regimes where coiled and stretched configurations are observed. Expectedly, we find that the transition to stretching occurs for large values of either $p_a$ (many dipoles) or $\sigma_0$ (stronger dipoles). We find that the transition is very well captured by a curve of constant $A=p_a\sigma_0$, with a critical value of $A_c\approx 4$ in the  case shown in Fig.~\ref{Fig:Fig6}(c).
The activity threshold $A_c$ is length-dependent, and is found to decay with $N$ as shown in Fig.~\ref{Fig:Fig6}(d): indeed, a longer chain will carry more dipoles at a fixed value of $p_a$. For intermediate chain lengths, we find $A_c\sim N^{-1}$,  {although this apparent scaling is only observed over one decade of $N$ (see inset in Fig.~\ref{Fig:Fig6}(d)).  A plateau is observed for longer chains, whose precise origin remains unclear; a possible explanation for this saturation may be that the nematic alignment resulting in stretching first occurs locally on small chain segments before the entire chain is able to unravel and stretch. These findings} suggest a transition governed by the parameter $AN$, which plays an analogous role as the Deborah number for the classic coil-stretch transition of flexible polymers in extensional flows \cite{BCAH1987}. Interestingly, we find that the transition to spontaneous stretching only occurs for $N\ge 4$, a  result that we explore further using a theoretical model in Sec.~\ref{sec:theory}.

\subsection{Theory for an active trimer\label{sec:theory}}

As observed in Fig.~\ref{Fig:Fig6}(d), $N=4$ is the lower limit above which the polymer undergoes the coil-stretch transition in our simulations. To understand this behavior, we further study the case of a trimer consisting of $N=3$ beads using a theoretical model based on a Fokker-Planck formulation. Denoting by $\mathbf{q}$ the generalized angular coordinates describing the configuration of the chain, the Fokker-Planck equation for the probability density function $\psi(\mathbf{q},t)$ is given by \cite{H1974,BCAH1987,PV2016}
\begin{align}
\begin{split}
\frac{\partial \psi}{\partial t}+\frac{\partial}{\partial q_i} \Bigg[\Gamma_{ij}  \Bigg(&\sum_{n=1}^N u^a_k(\mathbf{r}_n)\frac{\partial r_{n,k}}{\partial q_j} \psi  \\
&-\frac{k_\mathrm{B} T}{\zeta} \sqrt{h}\frac{\partial}{\partial q_j} \left(\frac{\psi}{\sqrt{h}}\right)\Bigg)\Bigg]=0, \label{Eq:FP}
\end{split}
\end{align}
where index $n$ refers to the bead number, and all other indices denote components of the various vector and tensor quantities, for which the Einstein summation convention applies. In Eq.~(\ref{Eq:FP}),  $\boldsymbol{\Gamma} = \mathbf{H}^{-1}$ and $h = \mathrm{det}(\mathbf{H})$ are defined in terms of the tensor
\begin{equation}
    H_{ij}=\sum_{n=1}^N \frac{\partial \mathbf{r}_n}{\partial q_i}\cdot \frac{\partial \mathbf{r}_n}{\partial q_j}.
\end{equation}

\begin{figure}[t]
    \centering
     \includegraphics[width=0.99\linewidth]{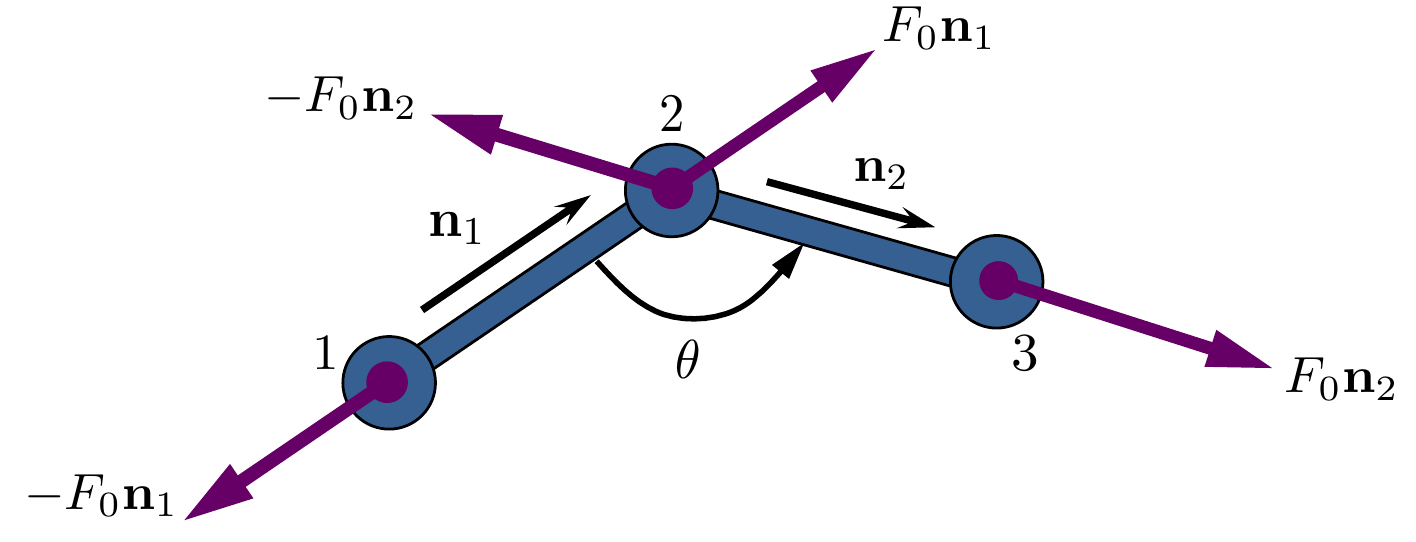}
     \hspace{0.0cm}
    \caption{Schematic of an active trimer, where the internal coordinate of interest is the bond angle $\theta=\pi-\cos^{-1}(\mathbf{n}_1\cdot\mathbf{n}_2)$.}
    \label{fig:fig8}
\end{figure}

In the case of interest where there is no external flow, the mean position and orientation of the chain simply undergo Brownian diffusion, and the only internal coordinate with a non-trivial probability distribution function is the bond angle $\theta=\pi-\cos^{-1}(\mathbf{n}_1\cdot\mathbf{n}_2)$ as depicted in Fig.~\ref{fig:fig8}. Eq.~(\ref{Eq:FP}) can be simplified in this case. Setting the orientational flux to zero at steady state yields the simple differential equation  
\begin{equation}
\ell^2 \dot{\theta}^a \psi -  \frac{k_B T}{\zeta} \sqrt{h} \frac{\mathrm{d}}{\mathrm{d} \theta}\Big(\frac{\psi}{\sqrt{h}}\Big) = 0, \label{Eq:Eq23}
\end{equation}
where $\dot{\theta}^a$ is the deterministic angular velocity resulting from the flow induced by the active dipoles, and $h=\ell^4(4-\cos^2\theta)/9$ \cite{PV2016}. Here, the probability density function is normalized as
\begin{equation}
    \int_{0}^\pi \psi(\theta) \sin\theta\,\mathrm{d}\theta=1, \label{eq:normalization}
\end{equation}
where the factor of $\sin\theta$ comes from the solid angle in three dimensions. 

For simplicity, we assume that both links are subject to permanent dipoles ($p_a=1$). The angular velocity in that case can be calculated as
\begin{equation}
    \dot{\theta}^a= \frac{1}{\sin\theta} \left(\dot{\mathbf{n}}_1\cdot\mathbf{n}_2+\mathbf{n}_1\cdot\dot{\mathbf{n}}_2\right) = \frac{2}{\sin\theta}\mathbf{n}_1\cdot\dot{\mathbf{n}}_2
\end{equation}
where
\begin{equation}
\dot{\mathbf{n}}_2=\frac{1}{\ell}(\mathbf{I}-\mathbf{n}_2\mathbf{n}_2)\cdot\mathbf{u}^a(\mathbf{r}_3).
\end{equation}
Upon substituting the expression for the active flow velocity at the location of bead 3,
\begin{equation}
    \mathbf{u}^a(\mathbf{r}_3)=\left[\mathbf{G}(\mathbf{r}_3;\mathbf{r}_2)-\mathbf{G}(\mathbf{r}_3;\mathbf{r}_1) \right]\cdot F_0\mathbf{n}_1,
\end{equation}
with $\mathbf{G}$ the Oseen tensor defined in Eq.~(\ref{eq:oseen}), we obtain after simplifications
\begin{equation}
    \dot{\theta}^a= \frac{F_0 }{4\pi\mu\ell^2}\sin\theta\left[1-\frac{3}{2\sqrt{2(1 - \cos \theta)}}\right].
\end{equation}
Inserting this expression into the flux balance Eq.~(\ref{Eq:Eq23}) then yields the governing equation for $\psi(\theta)$,
\begin{align}
\begin{split}
\frac{3}{2}\left(\frac{a}{\ell}\right) \sigma_0 \sin \theta&\left[1 - \frac{3}{2\sqrt{2(1 - \cos \theta)}} \right] \psi  = \\ 
    &\quad  \sqrt{4 - \cos^2 \theta}\frac{\mathrm{d}}{\mathrm{d} \theta}\left(\frac{\psi}{\sqrt{4 - \cos^2 \theta}}\right),
    \end{split}\label{Eq:Eq36} 
\end{align}
which is written in dimensionless form and involves both the dipole strength $\sigma_0$ and ratio $a/\ell$ of the bead hydrodynamic radius to the bond length.

\begin{figure}[t]
    \centering
    \includegraphics[width=0.9\linewidth]{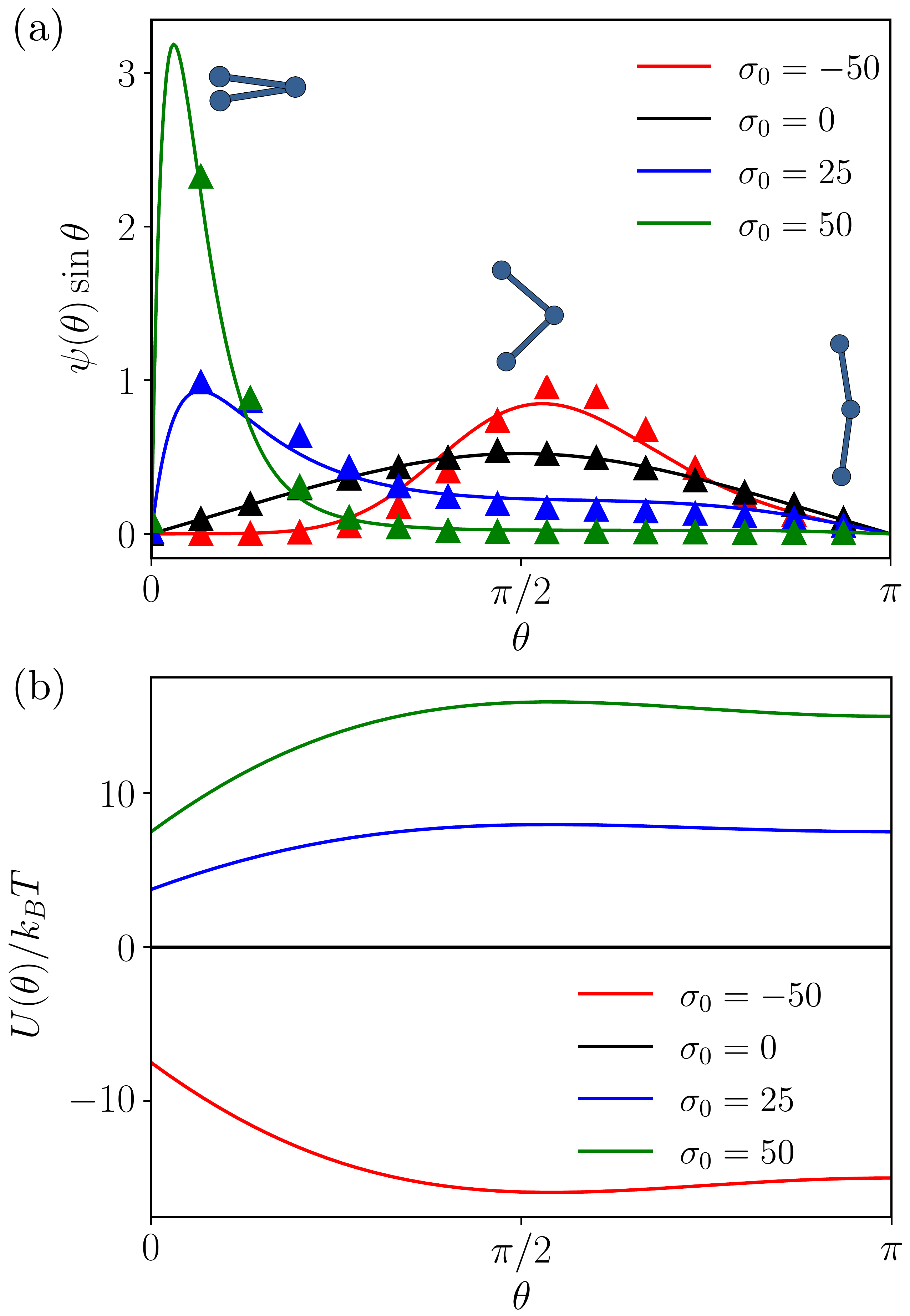}
    \caption{(a) Probability density function $\psi(\theta)\sin\theta$ of the bond angle $\theta$ for a trimer with different levels of activity and for $a/\ell=0.1$ (the factor of $\sin\theta$ arises from the solid angle in Eq.~(\ref{eq:normalization})).\ The plot compares the analytical solution of Eq.~(\ref{Eq:Eq36}) (solid lines) with results from Langevin simulations (symbols).\ (b) Effective potential energy $U(\theta)$ for active self-alignment as a function of bond angle $\theta$ (see Eq.~(\ref{eq:energy})).}
    \label{fig:fig9}
\end{figure}

In the passive case ($\sigma_0=0$), Eq.~(\ref{Eq:Eq36}) is readily integrated as
\begin{equation}
\psi_p(\theta)=\psi_0 \sqrt{4-\cos^2\theta} \qquad (\sigma_0=0),
\end{equation}
which is the classic solution for a rigid trimer at thermal equilibrium \cite{H1974}, which peaks at and is symmetric about $\theta=\pi/2$; see Fig.~\ref{fig:fig9}. The prefactor is obtained from the normalization condition Eq.~(\ref{eq:normalization}) as $\smash{\psi_0=(2\pi/3+\sqrt{3})^{-1}}$. When activity is present ($\sigma_0\neq 0$), the solution of Eq.~(\ref{Eq:Eq36}) for the distribution function becomes
\begin{equation}
    \psi_a(\theta)=\psi_p(\theta) \exp\left[-\frac{3a\sigma_0}{2\ell} \left(\cos\theta +\frac{3}{2}\sqrt{2(1-\cos\theta)}\right) \right], \label{eq:psia}
\end{equation}
with a new normalization constant 
$\psi_0$ that now depends on $(a/\ell)\sigma_0$ and must be determined numerically.\ We find that dipolar activity modifies the equilibrium distribution $\psi_p(\theta)$ through a Boltzmann factor of the form $\propto \exp(-U(\theta)/k_\mathrm{B} T)$ where 
\begin{equation}
    \frac{U(\theta)}{k_\mathrm{B} T}=\frac{3a\sigma_0}{2\ell} \left(\cos\theta +\frac{3}{2}\sqrt{2(1-\cos\theta)}\right) \label{eq:energy}
\end{equation}
can be interpreted as an effective potential energy for self-alignment of the chain under its own active flow. 

Figure~\ref{fig:fig9}(a) shows $\psi(\theta)\sin\theta$ as a function of $\theta$ for different values of $\sigma_0$, and compares the analytical solution of Eq.~(\ref{eq:psia}) with results from Langevin simulations. Good agreement is found between the two.\ As anticipated based on Fig.~\ref{Fig:Fig6}(d), we find that the behavior of the trimer is quite unlike that of longer chains: extensile trimers ($\sigma_0>0$) tend to remain in a folded configuration with $\theta\approx 0$, whereas contractile trimers prefer to be kinked with $\theta\approx \pi/2$. The propensity for the extensile trimer to remain folded can be understood by considering the potential energy in Fig.~\ref{fig:fig9}(b): in extensile cases with $\sigma_0>0$, the potential energy displays two minima at $\theta=0$ and $\pi$, which both correspond to configurations with aligned dipoles.\ However, the minimum at $\theta=0$ is significantly deeper, since the two dipoles are closer to one another and therefore interact more strongly in the folded state than in the unfolded state.\ This explains why the folded configuration is preferred in this case as seen in  Fig.~\ref{fig:fig9}(a).\ This folded state, however, becomes much less likely in longer chains, as the links also experience tension forces from the other parts of the chain. Excluded volume interactions also decrease the likelihood of folded conformations.

\section{Conclusions\label{sec:conclusions}}

We have used Langevin simulations to study the conformational dynamics of a model active polymer consisting of freely-jointed bead-rod chains decorated by stochastic force dipoles that drive long-ranged flows in the surrounding viscous solvent. Depending on the type of dipole, significantly different dynamics were observed: fluid flows driven by contractile dipoles tend to create transient kinks in the chain, resulting in coiled conformations with enhanced fluctuations; on the other hand, the flows driven by extensile dipoles tend to align neighboring chain segments, resulting in the spontaneous unfolding and stretching of the polymer above a critical level of activity. The dynamics in the extensile case are reminiscent of the coil-stretch transition of passive polymer chains in extensional flows \cite{D1974,BCAH1987}. However, the active coil-stretch transition uncovered here is self-induced and does not require any external flow or forcing. Upon unfolding and stretching of the polymer, the aligned dipoles along the chain drive a macroscopic fluid flow with extensional symmetry that resists relaxation under thermal fluctuations and sustains the stretched conformation along an arbitrary axis selected at random.  {Because the chain rigidifies under extensile activity, its configurational entropy is reduced and its behavior switches from a Gaussian chain to an effective wormlike chain.}  As demonstrated by our analysis, the transition to stretching occurs above a critical value of $N p_a \sigma_0$, which denotes the product of the mean number of dipoles $Np_a$ along the chain with the dimensionless dipole strength $\sigma_0$. While most of our simulations have neglected hydrodynamic interactions (other than those directly resulting from dipolar activity), we anticipate that a full characterization of the transition in the presence of hydrodynamic interactions may uncover yet richer behavior, such as conformation hysteresis near the transition point, which is known to occur in the case of the passive coil-stretch transition \cite{CBSC2003,HL2005}.

The active polymer model used in this work was motivated by interphase chromatin, which is a flexible polymer known to be subject to ATP-powered enzymes such as RNA polymerase \cite{zidovska2013micron}. A detailed microscopic model for the local active stresses generated by these enzymes is still lacking, yet we expect to be able to coarse-grain them as dipoles \cite{bruinsma2014chromatin}, and past work has suggested that fluid-mediated interactions between these dipoles may be responsible for the coherent motions observed inside the cell nucleus during interphase \cite{saintillan2018extensile}. Chromatin and its associated enzymes, however, cannot be extracted in vitro without major disruption to their structure and function, so that this system is hardly a good candidate for experimental validation of our model. Other synthetic systems, however, would be very well suited for that purpose. One example could consist of a flexible chain composed of bimetallic autophoretic rods \cite{PBKWMS2006}, which can be designed to drive either extensile or contractile flows \cite{BBLWRZWS2019}. Such a chain of extensile rods should display the active coil-stretch transition, which could be externally controlled by addition or removal of chemical fuel such as hydrogen peroxide to the solution. This suggests novel designs for smart polymeric materials whose effective  rheological, optical or thermal properties could be tuned reversibly by triggering the active coil-stretch transition on the microscale.

\begin{acknowledgments}
The authors thank M. J. Shelley and A. Zidovska for useful conversations, and gratefully acknowledge funding from National Science Foundation Grant CMMI-1762566. 
\end{acknowledgments}

\appendix
\section{Comparison with full hydrodynamics\label{sec:appendix}}

\begin{figure}[t]
    \centering
    \includegraphics[width=1\linewidth]{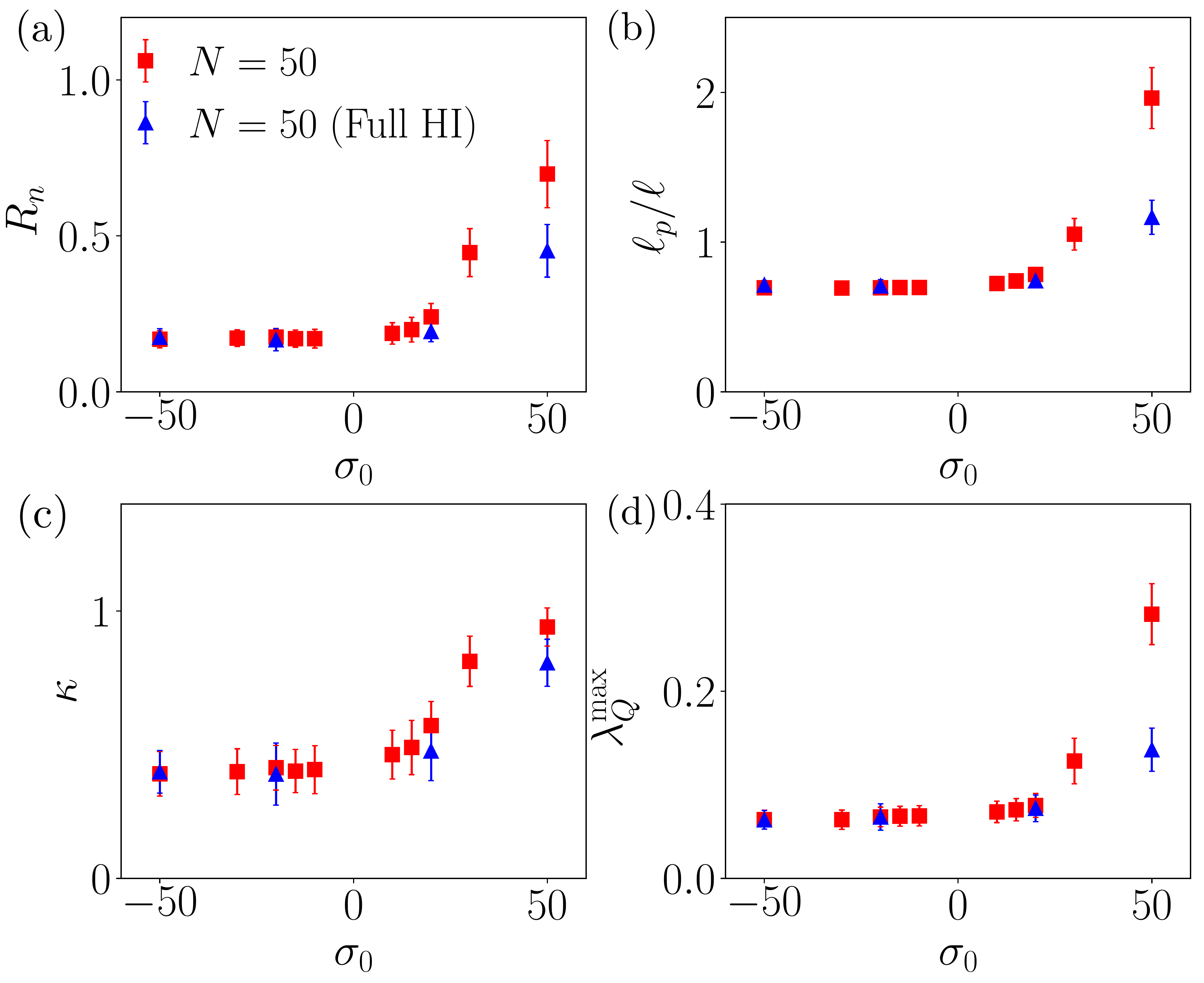}
    \caption{Effect of hydrodynamic interactions on the average steady-state values of (a) the scaled end-to-end distance $R_n$, (b) the effective persistence length $\ell_p/\ell$, (c) the relative shape anisotropy $\kappa$, and (d) the nematic scalar order parameter $\lambda_Q^\mathrm{max}$ as functions of $\sigma_0$ for chains of length $N=50$ with $k_{\mathrm{on}}=200$ and  $k_{\mathrm{off}}=500$. The plots compare results obtained using Eq.~(\ref{Eq:Eq1}), which neglects hydrodynamic interactions other than those induced by active dipoles, with Eq.~(\ref{Eq:HI}), which accounts for full hydrodynamic interactions. }\vspace{-0.3cm}
    \label{Fig:Fig10}
\end{figure}

The model discussed in Sec.~\ref{sec:model} describes a free-draining polymer chain where hydrodynamic interactions due to constraint forces, excluded volume interactions and thermal fluctuations are neglected and where the only hydrodynamic flow is that induced by active dipoles. Here, we present a more complete model that accounts for hydrodynamic interactions and compare results obtained by both formulations. In the presence of hydrodynamic interactions, the Langevin equation (\ref{Eq:Eq1}) becomes: \vspace{-0.2cm}
\begin{equation}
\frac{\mathrm{d}\mathbf{r}_i}{\mathrm{d}t}=\mathbf{u}^a(\mathbf{r}_i,t)+\sum_{j=1}^N\mathbf{M}_{ij}\cdot\left[\mathbf{F}^c_j(t)+\mathbf{F}^e_j(t)\right]+\boldsymbol{\xi}_i(t). \label{Eq:HI}
\end{equation}
 $\mathbf{M}_{ij}$ denotes the grand mobility tensor that captures viscous drag on the beads as well as long-ranged hydrodynamic interactions:
\begin{equation}
\mathbf{M}_{ij}=\frac{\mathbf{I}}{\zeta}\delta_{ij}+\mathbf{G}(\mathbf{r}_{i};\mathbf{r}_j)\,(1-\delta_{ij})\,, \label{Eq:Eq8a}
\end{equation}
where $\mathbf{G}$ is the Oseen tensor introduced in Eq.~(\ref{eq:oseen}). Furthermore, the fluctuation--dissipation theorem governing the statistics of Brownian displacements becomes:
\begin{equation}
 \langle \boldsymbol{\xi}_i(t)\rangle=\mathbf{0}, \quad \langle \boldsymbol{\xi}_i(t)\boldsymbol{\xi}_j(t')\rangle=2 k_\mathrm{B} T \mathbf{M}_{ij}\delta(t-t'). \label{Eq:Eq8}
\end{equation}
In practice, $\boldsymbol{\xi}_i$ is calculated as\vspace{-0.0cm}
\begin{equation}
    \boldsymbol{\xi}_i(t)=\sum_{j=1}^N \mathbf{B}_{ij}\cdot\mathbf{w}_j, \label{Eq:Eq8b}\vspace{-0.0cm}
\end{equation}
where $\mathbf{w}_j$ is an uncorrelated Gaussian white noise with zero mean and unit variance, and the tensor $\mathbf{B}_{ij}$ is related to the grand mobility tensor as\vspace{-0.1cm}
\begin{equation}
\sum_{p=1}^{N}\mathbf{B}_{ip}\cdot\mathbf{B}_{jp}^T=2 k_\mathrm{B} T \mathbf{M}_{ij}\,. \label{Eq:Eq9} 
\vspace*{0.2cm}\end{equation}
We compute $\mathbf{B}_{ij}$ as the lower triangular Cholesky factor of $\mathbf{M}_{ij}$.\ An additional complication in the presence of hydrodynamic interactions arises from the calculation of tensions: indeed,  {the left-hand side of} Eq.~(\ref{Eq:Eq14}) now involves a full matrix instead of a tridiagonal system, which we solve using LU decomposition.

Figure~\ref{Fig:Fig10} shows a comparison of conformational properties in simulations with and without hydrodynamic interactions.\ We find that all the qualitative trends reported in Sec.~\ref{sec:results} remain unchanged.  {We observe, however, that the transition to stretching in extensile systems is slightly delayed in the presence of hydrodynamic interactions as is especially visible in panels (a) and (c), and} that the spontaneous stretching induced by activity is weakened due to the increased viscous dissipation in the system.

\end{document}